# MODELING LONG-TERM LONGITUDINAL HIV DYNAMICS WITH APPLICATION TO AN AIDS CLINICAL STUDY

By Yangxin Huang[1] and Tao Lu

*University of South Florida*

A virologic marker, the number of HIV RNA copies or viral load, is currently used to evaluate antiretroviral (ARV) therapies in AIDS clinical trials. This marker can be used to assess the ARV potency of therapies, but is easily affected by drug exposures, drug resistance and other factors during the long-term treatment evaluation process. HIV dynamic studies have significantly contributed to the understanding of HIV pathogenesis and ARV treatment strategies. However, the models of these studies are used to quantify short-term HIV dynamics ($<1$ month), and are not applicable to describe long-term virological response to ARV treatment due to the difficulty of establishing a relationship of antiviral response with multiple treatment factors such as drug exposure and drug susceptibility during long-term treatment. Long-term therapy with ARV agents in HIV-infected patients often results in failure to suppress the viral load. Pharmacokinetics (PK), drug resistance and imperfect adherence to prescribed antiviral drugs are important factors explaining the resurgence of virus. To better understand the factors responsible for the virological failure, this paper develops the mechanism-based nonlinear differential equation models for characterizing long-term viral dynamics with ARV therapy. The models directly incorporate drug concentration, adherence and drug susceptibility into a function of treatment efficacy and, hence, fully integrate virologic, PK, drug adherence and resistance from an AIDS clinical trial into the analysis. A Bayesian nonlinear mixed-effects modeling approach in conjunction with the rescaled version of dynamic differential equations is investigated to estimate dynamic parameters and make inference. In addition, the correlations of baseline factors with estimated dynamic parameters are explored and some biologically meaningful correlation results are presented. Further, the estimated dynamic parameters in patients with virologic success were compared to those in patients with virologic failure and significantly

Received December 2006; revised May 2008.
[1]Supported in part by NIAID/NIH research Grant AI055290.

important findings were summarized. These results suggest that viral dynamic parameters may play an important role in understanding HIV pathogenesis, designing new treatment strategies for long-term care of AIDS patients.

**1. Introduction.** For the past decade HIV dynamics has been one of the most important areas in AIDS research [Ho et al. (1995), Perelson et al. (1996, 1997), Markowitz et al. (2003), Wu et al. (2005)]. It has since greatly improved our understanding of the pathogenesis of HIV-1 infection, and guided for the treatment of AIDS patients and evaluation of antiretroviral (ARV) therapies [Ding and Wu (2001), Wu et al. (2005)]. Analysis of the dynamics of HIV infection in response to drug therapy has elucidated crucial properties of viral dynamics. Long-term treatment in HIV infected patients with highly active antiretroviral therapies (HAART) results in a decrease of plasma HIV-1 RNA (viral load). The decay in viral load occurs in the first few weeks from beginning treatment [Perelson et al. (1997)]: it may be sustained for a long period, but often is followed by resurgence of plasma viral load within months [Nowak and May (2000)]. The resurgence of virus may be caused by drug resistance, poor patient adherence, pharmacokinetics and other factors during therapy [Bangsberg et al. (2000), Wahl and Nowak (2000)].

Figure 1(a) displays the early stage trajectories based on the first 35-day viral load data (in $\log_{10}$ scale) for 42 subjects enrolled in the AIDS clinical trial study–A5055 [Acosta et al. (2004)], Figure 1(b) includes the later stage ($> 35$ days) viral load data of the same patients. This study was a Phase I/II, randomized, open-label, 24-week comparative study of the pharmacokinetic, tolerability and ARV effects of two regimens of indinavir (IDV) and ritonavir (RTV), plus two nucleoside analogue reverse transcriptase inhibitors (NRTIs) on HIV-1-infected subjects failing protease inhibitor (PI)-containing ARV therapies [Acosta et al. (2004)]. 44 subjects who failed their first PI-containing regimens were randomized to one of two IDV/RTV regimens: IDV 800 mg twice daily (q12h) + RTV 200 mg q12h (Arm A) and IDV 400 mg q12h + RTV 400 mg q12h (Arm B). Out of the 44 subjects, 42 subjects are included in the analysis; for the remaining two subjects, one was excluded from the analysis since the pharmacokinetic (PK) parameters were not obtained and the other was excluded since PhenoSense HIV could not be completed on this subject due to an atypical genetic sequence that causes the viral genome to be cut by an enzyme used in the assay. Plasma HIV-1 RNA (viral load) measurements were taken at days 0, 7, 14, 28, 56, 84, 112, 140 and 168 of follow-up. The data for PK parameters ($C_{\min}$ which is the trough levels of drug concentration in plasma), phenotype marker (baseline and failure $IC_{50}$ which represents the drug concentration necessary to inhibit viral replication by 50% in vitro) and adherence from this study were also available. The adherence data were determined from pill-count data. A more detailed description of this study and data can be found in the publication by Acosta et al. (2004).

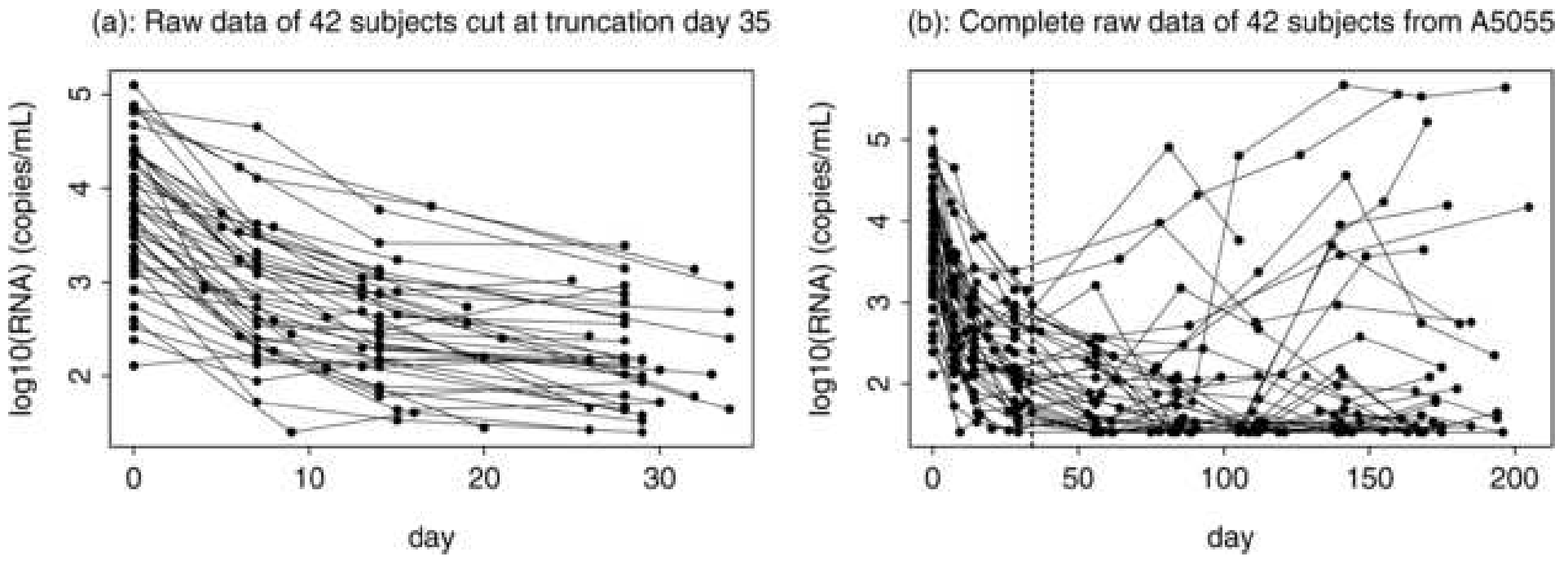

Fig. 1. *The profiles of viral load measured from RNA levels in plasma (in* $\log_{10}$ *scale) from a group of patients in an AIDS clinical study (A5055). Change in viral load during treatment is shown for day 0 to day 35* (a) *and to the end of study* (b).

Many HIV dynamic models have been proposed by AIDS researchers [Ho et al. (1995), Perelson et al. (1997, 1996), Wu and Ding (1999), Ding and Wu (2000), Nowak and May (2000), Huang, Rosenkranz and Wu (2003)] to provide theoretical principles in guiding the development of treatment strategies for HIV-infected patients. Unfortunately, these models are mostly developed to quantify short-term dynamics. While these models may reflect well short-term viral dynamics, they do not correctly describe long-term virologic responses with antiretroviral (ARV) therapies. In other words, these models fit only the early segment of the viral load trajectory [Figure 1(a)] and thus were limited to interpret typical HIV dynamic data resulting from AIDS clinical trials. Moreover, as is seen from Figure 1(b), the viral load trajectory may change to different shapes in the later stage due to, among other clinical factors, drug resistance and noncompliance. Although some studies [Jackson (1997), Labbé and Verttoa (2006), Verotta (2005)] explored long-term viral dynamic models with constant drug efficacy incorporating clinical factors such as drug adherence to describe virologic responses with ARV therapies, they considered viral dynamic models incorporating only one factor without accounting for other potential confounding factors. Huang and Wu (2006) extended the work of these studies to investigate dynamic models with time varying drug efficacy incorporating more than one clinical factor, but they only consider a single drug to quantify antiviral drug efficacy; in addition, the statistical model, in conjunction with the differential equations (2.2) in their study, is not parameter identifiable in the sense that different combinations of values of the nonidentifiable parameters can lead to the same likelihood, making it impossible to decide among the potential parameter values. Since HAART usually combines two or more ARV drugs, there is a need to develop models with a priori identifiable parameters that can describe long-term viral dynamics with more than one drug treatment effect. Thus, practical models and effective statistical modeling methods must address the following technical issues: (i) how to obtain a model with a priori identifiable parameters (from the perspective of the likelihood); (ii) how to account for the confounding factor effects of drug concentration, susceptibility and adherence on antiviral responses with ARV therapies of more than one drug combination treatment; (iii) how to investigate associate methodologies in conjunction with models specified by a system of differential equations with a time varying coefficient, but without a closed-form solution. Such extensive research efforts are warranted to study the intermediate to long-term drug effects on HIV through the modeling of virologic markers.

Data from viral dynamic studies usually consist of repeated viral load measurements taken over treatment time course for each subject. The intra-individual variation and the inter-individual variation are usually modeled by a two-stage hierarchical model. The first stage specifies the mean and covariance structure for a given individual, while the second stage characterizes the inter-individual variation. Such models are often referred to as nonlinear mixed-effects (NLME) models. Understanding the nature of inter-individual systemic and random variation at the second stage often receives far more emphasis. As is evident from Figure 1, the inter-patient variations appear to be large, in particular, for the long-term period. Much of this inter-individual variation may be explained by clinical factors such as drug exposures and drug resistance. Because the viral dynamic processes share certain similar patterns between patients while still having distinct individual characteristics, the hierarchical NLME models are often used to quantify individual heterogeneity among subjects [Wu, Ding and De Gruttola (1998), Wu and Ding (1999)]. Although the NLME model fitting can be implemented in standard statistical software, such as the function `nlme()` in S-plus [Pinheiro and Bates (2000)] the procedure `NLMIXED` in SAS (2000), it is difficult to use these standard packages in fitting NLME models when the closed form of the nonlinear function is not available.

Viral dynamic models can be formulated through differential equations. But there has been only limited development of statistical methodologies for estimating such models or assessing their agreement with observed data. The purpose of this paper is to develop long-term viral dynamic models based on a system of differential equations with time-varying coefficients but without closed-form solutions. It also aims to investigate associated statistical methodologies in conjunction with the viral dynamic models with application to an AIDS clinical trial study. Our dynamic model should be able to characterize sustained suppression or resurgence of the virus as arising from intrinsic viral dynamics, and/or influenced by clinical factors such as pharmacokinetics, compliance to treatment and drug susceptibility. We combine the Bayesian approach and mixed-effects modeling method to estimate both population and individual dynamic parameters under a framework of the hierarchical Bayesian NLME model. The paper is organized as follows. In Section 2 we introduce the various HIV dynamic systems and propose a simplified and rescaled viral dynamic model with time-varying drug efficacy which incorporates the effects of PK variation, drug resistance and adherence. In Section 3 a Bayesian approach implemented using the Markov chain Monte Carlo (MCMC) techniques is employed to estimate dynamic parameters for inference. The proposed methodology is applied to the data for pharmacokinetics, drug resistance and adherence, as well as the viral load from the AIDS clinical trial described in Section 1, and the results are presented in Section 4. Finally, the paper concludes with some discussions in Section 5.

## 2. HIV dynamic models.

2.1. *Structural models with constant drug efficacy.* The structural model is a mathematical description of HIV dynamics not including covariates and, in particular, drug exposure and resistance to treatment in the description of the data. Basic models of viral dynamics describe the interaction between cells susceptible to target cells ($T$), infected cells ($T^*$) and free virus ($V$). The most common model with imperfect drug effect is expressed in terms of the following set of differential equations [Perelson and Nelson (1999)]

$$\begin{aligned} \frac{d}{dt}T &= \lambda - d_T T - [1-\gamma_0]kTV, \\ \frac{d}{dt}T^* &= [1-\gamma_0]kTV - \delta T^*, \\ \frac{d}{dt}V &= N\delta T^* - cV, \end{aligned} \tag{2.1}$$

where $\lambda$ represents the rate at which new $T$ cells are created from sources within the body, such as the thymus, $d_T$ is the death rate of $T$ cells, $k$ is the infection rate of $T$ cells infected by virus, $\delta$ is the death rate for infected cells, $N$ is the number of new virions produced from each of the infected cells during their life-time, and $c$ is the clearance rate of free virions. The parameter $\gamma_0$ $(0 \le \gamma_0 \le 1)$ denotes the constant antiviral drug efficacy. If the regimen is not 100% effective ($\gamma_0 \neq 1$), the system of ordinary differential equations can not be solved analytically. The solutions to (2.1) then have to be evaluated numerically. When $\gamma_0 = 0$ (the drug has no effect), the model (2.1) reduces to the model in the publications [Nowak et al. (1995), Nowak and May (2000), Stafford et al. (2000)] while $\gamma_0 = 1$ (the drug is 100% effective), the model (2.1) reverts to the model discussed by Perelson et al. (1996) and Nowak and May (2000).

2.2. *Viral dynamic model with time-varying drug efficacy.* Since the model (2.1) assumes drug efficacy with a constant and does not consider the fact of variability in drug susceptibility (drug resistance) and adherence in the presence of ARV therapy, this model is only used to quantify short-term dynamics and does not correctly describe the long-term virologic responses with ARV treatment. In other words, the model is only used to fit the early segment data of the viral load trajectory. By considering drug exposures and drug susceptibility, a long-term HIV dynamic model can be expressed as

$$\begin{aligned} \frac{d}{dt}T &= \lambda - d_T T - [1-\gamma(t)]kTV, \\ \frac{d}{dt}T^* &= [1-\gamma(t)]kTV - \delta T^*, \\ \frac{d}{dt}V &= N\delta T^* - cV. \end{aligned} \tag{2.2}$$

Although this model also includes the interaction of target uninfected cells, infected cells that actively produce viruses and free virus, it differs from the previous model (2.1) in that the model (2.2) includes a time-varying parameter $\gamma(t)$ (as defined below), which quantifies the time-varying antiviral drug efficacy.

If we assume that the system of equations (2.2) is in a steady-state before initiating ARV treatment, then it is easy to show that the initial conditions for the system are

$$T_0 = \frac{c}{kN}, \qquad T_0^* = \frac{cV_0}{\delta N}, \qquad V_0 = \frac{\lambda N}{c} - \frac{d_T}{k}. \tag{2.3}$$

As Huang, Rosenkranz and Wu (2003) have shown, there exists a drug efficacy threshold $e_c = 1 - cd_T/kN\lambda$ such that if $\gamma(t) > e_c$ for all $t$, the virus will be eventually eradicated in theory. However, if $\gamma(t) < e_c$ (treatment is not potent enough) or if the potency falls below $e_c$ before virus eradication (due to drug resistance, e.g.), viral load may rebound [see Huang, Rosenkranz and Wu (2003) for a detailed discussion]. We briefly discuss the drug efficacy model with two or more agents which includes the model of one agent described in Huang and Wu (2006) as a special case.

Within the population of HIV virions in a human host, there is likely to be genetic diversity and corresponding diversity in sensitivity to the various ARV agents. In clinical practice, genotypic or phenotypic tests can be performed to determine the sensitivity of HIV-1 to ARV agents before a treatment regimen is selected. Here we use the phenotypic marker, the median inhibitory concentration ($IC_{50}$) [Molla et al. (1996)] to quantify agent-specific drug susceptibility. To model within-host changes over time in $IC_{50}$

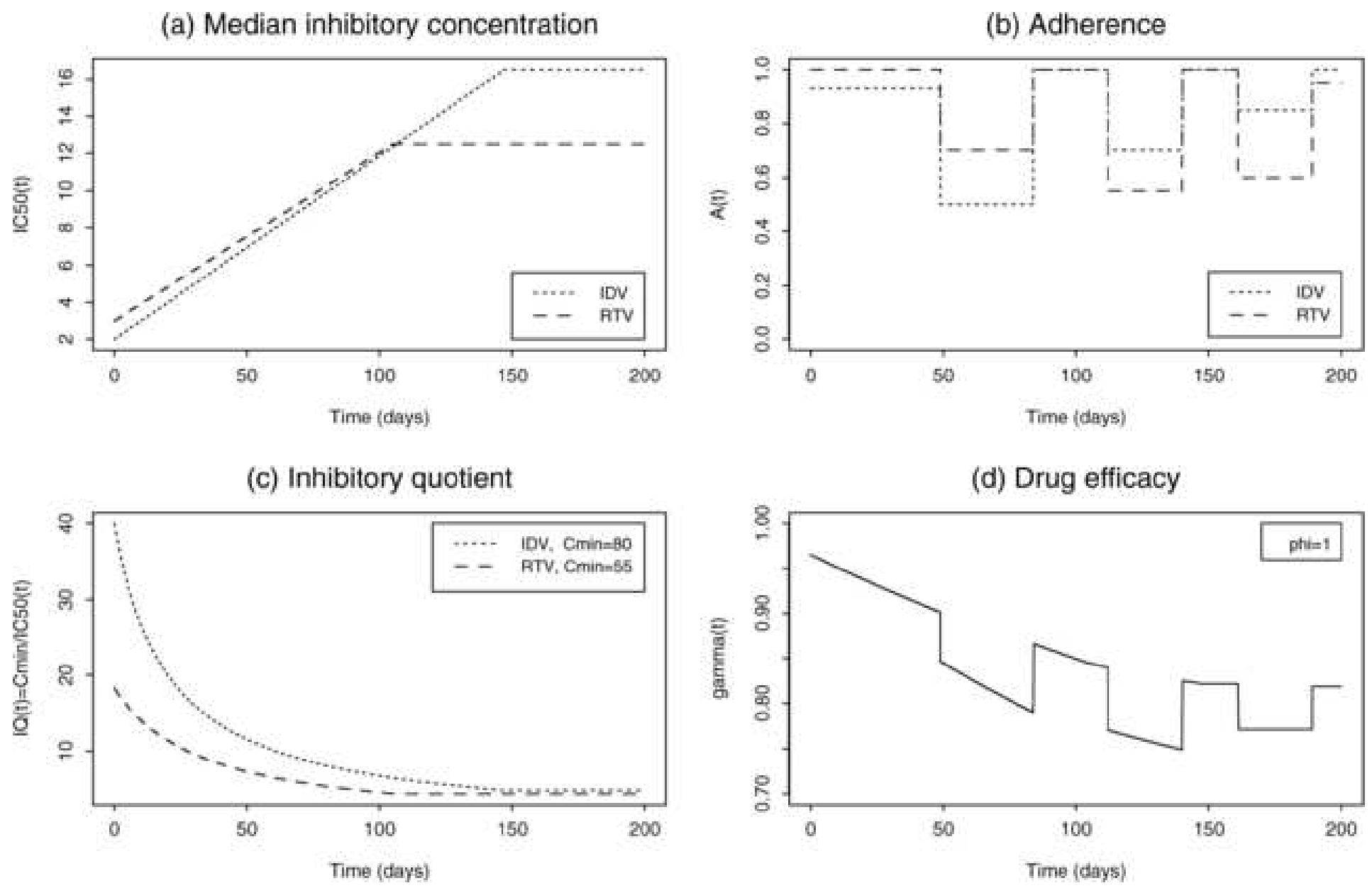

FIG. 2. (a) *The median inhibitory concentration curve* $[IC_{50}(t)]$*;* (b) *the time-course of adherence* $[A(t)]$*;* (c) *the time-course of inhibitory quotient* $[IQ(t)]$*;* (d) *the time-course of antiviral drug efficacy* $[\gamma(t)]$*.*

due to the emergence of new drug resistant mutations, the following function can be used:

$$IC_{50}(t) = \begin{cases} I_0 + \dfrac{I_r - I_0}{t_r} t, & \text{for } 0 < t < t_r, \\ I_r, & \text{for } t \geq t_r, \end{cases} \tag{2.4}$$

where $I_0$ and $I_r$ are respective values of $IC_{50}(t)$ at baseline and time point $t_r$ at which the resistant mutations dominate. In our study $t_r$ is the time of virologic failure. For subjects without a failure time $IC_{50}$, baseline $IC_{50}$ was held constant overtime. In other words, if $I_r = I_0$, no new drug resistant mutation is developed during treatment. Although more complicated models for median inhibitory concentration have been proposed based on the frequencies of resistant mutations and cross-resistance patterns [Wainberg et al. (1996), Bonhoeffer, Lipsitch and Levin (1997)] in clinical studies or clinical practice it is common to collect $IC_{50}$ values only at baseline and failure time as designed in A5055. Thus, this function may serve as a good approximation. As an example, such function for two ARV drugs is plotted in Figure 2(a).

Poor adherence to a treatment regimen is one of the major causes of treatment failure [Besch (1995), Ickovics and Meisler (1997)]. For various reasons patients may occasionally miss doses, may misunderstand prescription instructions or may miss multiple consecutive doses. These deviations from prescribed dosing affect drug exposure in predictable ways. We use the following model to represent adherence:

$$A_d(t) = \begin{cases} 1, & \text{for } T_k < t \le T_{k+1}, \text{ if all doses are taken in } [T_k, T_{k+1}], \\ R_d, & \text{for } T_k < t \le T_{k+1}, \text{ if } 100R_d\% \\ & \text{doses are taken in } [T_k, T_{k+1}], \end{cases} \tag{2.5}$$

where $0 \le R_d < 1$ $(d = 1, 2)$, with $R_d$ indicating the adherence rate for drug $d$ (in our study we focus on the two PI drugs of the prescribed regimen). $T_k$ denotes the adherence evaluation time at the $k$th clinical visit. As an example, Figure 2(b) shows the effect of adherence over time for two ARV drugs, respectively.

The HAART, containing two or more nucleoside/nonnucleoside reverse transcriptase inhibitors (RTI) and protease inhibitors (PI), has proven to be effective in reducing the amount of virus in the blood and tissues of HIV-infected patients. To model the relationship of drug exposure and resistance with antiviral efficacy, we employ the following modified $E_{\max}$ model [Sheiner (1985)] to represent the time-varying drug efficacy for two ARV agents within a class (e.g., the two PI drugs IDV and RTV):

$$\gamma(t) = \frac{IQ_1(t)A_1(t) + IQ_2(t)A_2(t)}{\phi + IQ_1(t)A_1(t) + IQ_2(t)A_2(t)}, \tag{2.6}$$

where $IQ_d(t) = C^d_{\min}/IC^d_{50}(t)$ $(d = 1, 2)$ denotes the inhibitory quotient (IQ) [Hsu et al. (2000)], $C^d_{\min}$, $A_d(t)$ and $IC^d_{50}(t)$ $(d = 1, 2)$ are the trough levels of drug concentration in plasma, adherence profile and the median inhibitory concentrations for the two agents, respectively. Note that $C_{\min}$ could be replaced by other PK parameters such as the area under the plasma concentration-time curve (AUC). Although $IC_{50}(t)$ can be measured by phenotype assays in vitro, it may not be equivalent to the $IC_{50}(t)$ in vivo. The parameter $\phi$ is used to quantify the conversion between in vitro and in vivo $IC_{50}$ that can be estimated from clinical data. $\gamma(t)$ ranges from 0 to 1, implying that one drug appears to be equally effective as the other. If $\gamma(t) = 1$, the drug is 100% effective, whereas if $\gamma(t) = 0$, the drug has no effect. Note that, if $C^d_{\min}$, $A_d(t)$ and $IC^d_{50}(t)$ are measured from a clinical study and $\phi$ can be estimated from clinical data, then the time-varying drug efficacy $\gamma(t)$ can be estimated for the whole period of ARV treatment. Lack of adherence reduces the drug exposure, which can be quantified by equation (2.5), and thus, based on the formula (2.6), reduces the drug efficacy which, in turn, can affect virologic response. The examples of the time courses of the inhibitory quotients and the drug efficacy $\gamma(t)$ with $\phi = 1$, $C^1_{\min} = 80$ and $C^2_{\min} = 50$ for two PI drugs are shown in Figures 2(c) and (d), respectively.

2.3. *Model reparametrization.* It is challenging to estimate all the seven parameters in the model (2.2) and conduct inference because this model is not a priori identifiable [in the sense of Cobelli, Lepschy and Jacur (1997): multiple sets of parameters obtain identical fits to the data], given only viral load measurements. To obtain a model with a priori identifiable parameters, the model (2.2) is reparametrized using the rescaled variables $\widetilde{T} = (d_T/\lambda)T$, $\widetilde{T}^* = (\delta/\lambda)T^*$, $\widetilde{V} = (k/d_T)V$. These yield the rescaled version as follows:

$$
\begin{aligned}
&\frac{d}{dt}\widetilde{T} = d_T(1 - \widetilde{T} - [1 - \gamma(t)]\widetilde{T}\widetilde{V}),\\
&\frac{d}{dt}\widetilde{T}^* = \delta([1 - \gamma(t)]\widetilde{T}\widetilde{V} - \widetilde{T}^*), \qquad (2.7)\\
&\frac{d}{dt}\widetilde{V} = c(R_0\widetilde{T}^* - \widetilde{V}),
\end{aligned}
$$

where $R_0 = kN\lambda/(cd_T)$ represents the basic reproductive ratio for the virus, defined as the number of newly infected cells that arise from any one infected cell when almost all cells are uninfected [Nowak and May (2000), Verotta (2005)]. Note that the rescaled model (2.7) has fewer parameters than the "original" model (2.2). We reparameterize the dynamic model so that identifiability of model (2.7) is guaranteed [Cobelli, Lepschy and Jacur (1997), Verotta (2005)] and parameters of the model can be uniquely identified. For model (2.7), the drug efficacy threshold $e_c = 1 - 1/R_0$.

If $R_0 < 1$, then the virus will not spread, since every infected cell will on average produce less than one infected cell. If, on the other hand, $R_0 > 1$, then every infected cell will on average produce more than one newly infected cell and the virus will proliferate. For the HIV to persist in the host, infected cells must produce at least one secondary infection, and $R_0$ must be greater than unity. Moreover, the greater the magnitude of $R_0$ is, the greater the degree of infection would be. If we assume that on a scale of weeks or days, the $T$-cell counts, infected cell counts and free virus do not change before the beginning of drug therapy, then we can compute a full pretreatment steady state [Perelson and Nelson (1999), Verotta (2005)]. Thus, initial conditions for the model can now be expressed as simple functions of the initial condition of viral load ($\widetilde{V}_0$): $\widetilde{T}_0 = 1/(1 + \widetilde{V}_0), \widetilde{T}_0^* = \widetilde{V}_0/(1 + \widetilde{V}_0), \widetilde{V}_0 = \widetilde{V}(0)$. The assumption of the initial steady state is necessary to guarantee identifiable (none of the models reported or referenced here is identifiable if the initial states are unknown), and is often justified by the clinical trial protocol [Nowak and May (2000), Verotta (2005)]. For example, in ACTG protocol 5055 (A5055), individual patients were taken off the drug before the initiation of the new therapy (washout period) to eliminate the effect of previously administered drugs and to guarantee that all individuals started from steady-state conditions. Finally, the predicted viral load ($V_p$) related to an equation output of viral load amount ($\widetilde{V}$) in model (2.7), such as $V_p = \rho\widetilde{V}$, is required to complete the rescaled model, where $\rho$, which is the equivalent of a volume of distribution for pharmacokinetics, is a viral load scaling (proportionality) factor (10,000 copies/mL) to be estimated from the data; $\widetilde{V}$ is generally expressed in copies RNA per unit volume.

**3. Bayesian nonlinear mixed-effects models.** A Bayesian nonlinear mixed-effects (BNLME) model allows us to incorporate prior information at the population level into the estimates of dynamic parameters for individual subjects. Denote the number of subjects by $n$ and the number of measurements on the ith subject by $m_i$. For notational convenience, let $\boldsymbol{\mu} = (\log\phi, \log c, \log\delta, \log d_T, \log\rho, \log R_0)^T$, $\boldsymbol{\theta}_i = (\log\phi_i, \log c_i, \log\delta_i, \log d_{Ti}, \log\rho_i, \log R_{0i})^T$, $\boldsymbol{\Theta} = \{\boldsymbol{\theta}_i, i = 1, \ldots, n\}$, $\boldsymbol{\Theta}_{\{i\}} = \{\boldsymbol{\theta}_l, l \neq i\}$ and $\mathbf{Y} = \{y_{ij}, i = 1, \ldots, n; j = 1, \ldots, m_i\}$. Let $f_{ij}(\boldsymbol{\theta}_i, t_j) = \log_{10}(V_p(\boldsymbol{\theta}_i, t_j))$, where $V_p(\boldsymbol{\theta}_i, t_j)$ is proportional to the numerical solution of $\widetilde{V}(t)$ in the differential equations (2.7) for the $i$th subject at time $t_j$. Let $y_{ij}(t)$ and $e_i(t_j)$ denote the repeated measurements of the common logarithmic viral load and a measurement error with mean zero, respectively. Note that log-transformation of dynamic parameters and viral load is used to make sure estimates of dynamic parameters are positive and to stabilize the variance, respectively. The BNLME model can be written in the following three stages [Gelfand et al. (1990), Davidian and Giltinan (1995), Wakefield (1996)].

*Stage* 1. Within-subject variation:

$$\mathbf{y}_i = \mathbf{f}_i(\boldsymbol{\theta}_i) + \mathbf{e}_i, \qquad \mathbf{e}_i|\sigma^2, \boldsymbol{\theta}_i \sim \mathcal{N}(\mathbf{0}, \sigma^2\mathbf{I}_{m_i}), \tag{3.1}$$

where $\mathbf{y}_i = (y_{i1}(t_1), \ldots, y_{im_i}(t_{m_i}))^T$, $\mathbf{f}_i(\boldsymbol{\theta}_i) = (f_{i1}(\boldsymbol{\theta}_i, t_1), \ldots, f_{im_i}(\boldsymbol{\theta}_i, t_{m_i}))^T$, $\mathbf{e}_i = (e_i(t_1), \ldots, e_i(t_{m_i}))^T$.

*Stage* 2. Between-subject variation:

$$\boldsymbol{\theta}_i = \boldsymbol{\mu} + \mathbf{b}_i, \qquad [\mathbf{b}_i|\boldsymbol{\Sigma}] \sim \mathcal{N}(\mathbf{0}, \boldsymbol{\Sigma}). \tag{3.2}$$

*Stage* 3. Hyperprior distributions:

$$\sigma^{-2} \sim Ga(a, b), \qquad \boldsymbol{\mu} \sim \mathcal{N}(\boldsymbol{\eta}, \boldsymbol{\Lambda}), \qquad \boldsymbol{\Sigma}^{-1} \sim Wi(\boldsymbol{\Omega}, \nu), \tag{3.3}$$

where the mutually independent Gamma ($Ga$), Normal ($\mathcal{N}$) and Wishart ($Wi$) prior distributions are chosen to facilitate computations [Davidian and Giltinan (1995)]. In this modeling analysis, the data are assumed to be missing at random (MAR) and, thus, analysis will give valid inferences [Heitjan and Basu (1996)].

Following the studies by Gelfand and Smith (1990) and Davidian and Giltinan (1995), it is shown from (3.1)–(3.3) that the full conditional distributions for the parameters $\sigma^{-2}, \boldsymbol{\mu}$ and $\boldsymbol{\Sigma}^{-1}$ may be written explicitly as

$$[\sigma^{-2}|\boldsymbol{\mu}, \boldsymbol{\Sigma}^{-1}, \boldsymbol{\Theta}, \mathbf{Y}]$$

$$\sim Ga\left(a+\frac{\sum_{i=1}^{n} m_i}{2}, \left\{\frac{1}{b}+\frac{1}{2}\sum_{i=1}^{n}\sum_{j=1}^{m_i}[y_{ij}-f_{ij}(\boldsymbol{\theta}_i,t_j)]^2\right\}^{-1}\right), \tag{3.4}$$

$$[\boldsymbol{\mu}|\sigma^{-2},\boldsymbol{\Sigma}^{-1},\boldsymbol{\Theta},\mathbf{Y}] \sim \mathcal{N}\left((n\boldsymbol{\Sigma}^{-1}+\boldsymbol{\Lambda}^{-1})^{-1}\left(\boldsymbol{\Sigma}^{-1}\sum_{i=1}^{n}\boldsymbol{\theta}_i+\boldsymbol{\Lambda}^{-1}\boldsymbol{\eta}\right), (n\boldsymbol{\Sigma}^{-1}+\boldsymbol{\Lambda}^{-1})^{-1}\right), \tag{3.5}$$

$$[\boldsymbol{\Sigma}^{-1}|\sigma^{-2},\boldsymbol{\mu},\boldsymbol{\Theta},\mathbf{Y}] \sim Wi\left(\left[\boldsymbol{\Omega}^{-1}+\sum_{i=1}^{n}(\boldsymbol{\theta}_i-\boldsymbol{\mu})(\boldsymbol{\theta}_i-\boldsymbol{\mu})^T\right]^{-1}, n+\nu\right). \tag{3.6}$$

Here, however, the full conditional distribution of each $\boldsymbol{\theta}_i$, given the remaining parameters and the data, cannot be calculated explicitly. The distribution of $[\boldsymbol{\theta}_i|\sigma^{-2},\boldsymbol{\mu},\boldsymbol{\Sigma}^{-1},\boldsymbol{\Theta}_{\{i\}},\mathbf{Y}]$ has a density function which is proportional to

$$\exp\left\{\frac{-\sigma^{-2}}{2}\sum_{j=1}^{m_i}[y_{ij}-f_{ij}(\boldsymbol{\theta}_i,t_j)]^2-\frac{1}{2}(\boldsymbol{\theta}_i-\boldsymbol{\mu})^T\boldsymbol{\Sigma}^{-1}(\boldsymbol{\theta}_i-\boldsymbol{\mu})\right\}. \tag{3.7}$$

To carry out the Bayesian inference, we need to specify the values of the hyper-parameters in the prior distributions. In the Bayesian approach we only need to specify the priors at the population level which are easy to obtain from previous studies or reference literature and usually are accurate and reliable. The values of hyper-parameters in this study were determined from the previous publications [Ho et al. (1995), Perelson et al. (1996), Perelson et al. (1997), Perelson and Nelson (1999), Nowak and May (2000), Verotta (2005)]. After we specify the model for the observed data and the prior distributions for the unknown parameters, we can draw statistical inference for the unknown parameters based on their posterior distributions. In the above Bayesian modeling approach evaluation of the required posterior distributions in a closed-form solution is prohibitive. However, as indicated above, it is relatively straightforward to derive either full conditional distributions for some parameters or explicit expressions which are proportional to the corresponding full conditional distributions for other parameters.

The MCMC scheme for drawing samples from the posterior distributions of all parameters in the above three stage model is obtained by iterating between the following two steps: (i) sampling from one of the conditional distributions (3.4)–(3.6); (ii) sampling from the expression (3.7). To implement an MCMC algorithm, here the Gibbs sampler is used to update $\sigma^{-2}, \boldsymbol{\mu}$ and $\Sigma^{-1}$, while we update $\boldsymbol{\theta}_i$ $(i = 1, 2, \ldots, n)$ using the Metropolis–Hastings (M–H) algorithm. After collecting the final MCMC samples, we are able to draw statistical inference for the unknown parameters. In particular, we are interested in the posterior means and quantiles. See the articles [Carlin and Louis (1996), Gelfand et al. (1990), Huang and Wu (2006), Roberts (1996) and Wakefield (1996)] for detailed discussions of the Bayesian modeling approach and the implementation of the MCMC procedures, including the choice of the hyper-parameters, the iterative MCMC algorithm, the choice of proposal density related to M–H sampling, sensitivity analysis and convergence diagnostics.

The hierarchical Bayesian approach allows us not only to borrow information from previous studies, but also to borrow information across-patients in the same study to estimate the dynamic parameters for an individual patient. Across-patient information is incorporated in Stages 1 and 2 via Bayesian theories, and the prior information regarding the estimates of the viral dynamic parameters from previous studies is incorporated in Stage 3 in the hierarchical model (3.1)–(3.3). Thus, the estimates of viral dynamic parameters for an individual patient are based on the data from this particular patient, the data from other patients in the same study and the prior information from previous studies. The information from these three resources is weighted according to the uncertainty of each information component in an efficient and optimal way via Bayesian theories. In contrast, the nonlinear least squares (NLS) method fits a model to the data from individual patients at a time. The data from an individual patient sometimes may not be enough to reliably identify all the viral dynamic parameters, and hence, data from other patients are generally not used. Thus, the key advantage of the hierarchical Bayesian approach, compared to the NLS method, is its efficient utilization of all the information available at hand.

For individual dynamic estimates, we may carry out further analysis for the estimated dynamic parameters using standard statistical methods [Ding and Wu (2001)]. We correlated the estimated viral dynamic parameters with baseline host factors (baseline viral load, CD4, age and weight) and summary statistics of virologic/immunologic responses. We used the Spearman rank test to evaluate statistical significance of correlations. Simple linear relationships were explored using a robust linear regression method (MM-estimator) due to the presence of outliers (although some of the relationships may be nonlinear) [Venables and Ripley (1999)]. All $P$-values are two-sided with a significance level of 0.05, and no adjustments for multiple testing were made in this analysis.

**4. Analysis of A5055 data.** In this section we apply the BNLME modeling approach for fitting the A5055 data described in Section 1. Based on the discussion in Section 3, the prior distribution for $\boldsymbol{\mu}$ was assumed to be $\mathcal{N}(\boldsymbol{\eta}, \boldsymbol{\Lambda})$, with $\boldsymbol{\Lambda}$ being a diagonal matrix. The details of the prior construction for unknown parameters are discussed in Huang and Wu (2006). Thus, the values of the hyper-parameters are chosen as follows [Ho et al. (1995), Perelson et al. (1996, 1997), Perelson and Nelson (1999), Nowak and May (2000), Verotta (2005)]:

$$a = 4.5, \qquad b = 9.0, \qquad \nu = 8.0,$$
$$\boldsymbol{\eta} = (4.0, 1.1, -1.0, -2.5, 1.4, 0.28)^T,$$
$$\boldsymbol{\Lambda} = \mathrm{diag}(1000.0, 1000.0, 1000.0, 1000.0, 1000.0, 1000.0),$$
$$\boldsymbol{\Omega} = \mathrm{diag}(2.0, 2.0, 2.0, 2.0, 2.0, 2.0).$$

Note that the noninformative priors are chosen for all the parameters. As suggested by Geman and Geman (1984), for example, one long run may be more efficient with considerations of the following two points: (i) a number of initial "burn-in" simulations are discarded, since from an arbitrary starting point it would be unlikely that the initial simulations came from the stationary distribution targeted by the Markov chain; (ii) one may only save every $k$th ($k$ being an integer) simulation samples to reduce the dependence among samples used for parameter estimation. We are going to adopt these strategies in our MCMC implementation using FORTRAN code that calls a differential equation subroutine solver (DIVPRK) in IMSL library (1994), which uses the Runge–Kutta–Verner fifth-order method. The computer codes are available from the corresponding author upon request. An informal check of convergence is conducted based on graphical techniques according to the suggestion of Gelfand and Smith (1990). Based on the results, we propose that, after an initial number of 30,000 burn-in iterations, every 5th MCMC sample was retained from the next 120,000 samples. Thus, we obtained 24,000 samples of targeted posterior distributions of the unknown parameters.

4.1. *Model fitting and parameter estimation results.* The dynamic model (2.7) was fitted to the viral load data from 42 patients using the proposed Bayesian approach. We report the individual dynamic parameter estimates in Figure 3 and their summary statistics in Table 1. We observed a large between-subject variation in the estimates of all individual dynamic parameters [the coefficient of variation (CV) ranges from 18.2% to 175.0% for different dynamic parameters]. For instance, the smallest virus clearance rate ($c$) was estimated as 2.157 day$^{-1}$ with a corresponding half-life of 0.32 ($= \log 2/c$) days or 7.7 hours, and the largest was 6.448 day$^{-1}$ with a half-life of 0.1 days or 2.4 hours. The smallest death rate of infected cells ($\delta$) was estimated as 0.08 day$^{-1}$ with a corresponding half-life of 8.7 days and the largest was estimated as 0.641 day$^{-1}$ with a half-life of 1 day.

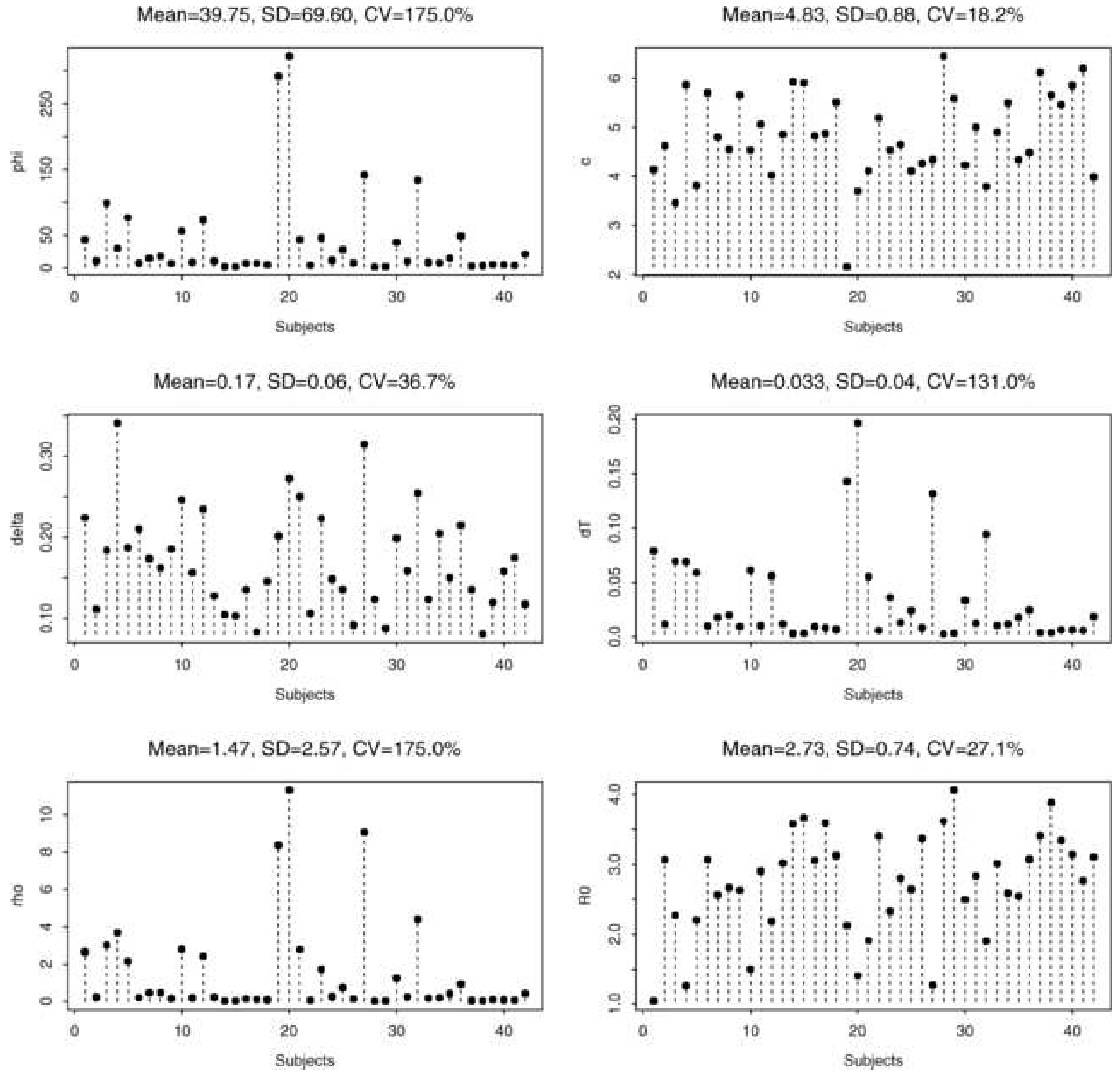

FIG. 3. *The estimated parameters of subject-specific individuals based on an AIDS clinical trial dataset with 42 patients. SD and CV=SD/Mean denote the standard deviation and coefficient of variation, respectively.*

The population posterior means and the corresponding 95% equal-tail credible intervals (CI) for the six parameters are summarized in Table 2. It is shown that the population estimates are 4.736 and 0.360 for $c$ and $\delta$, respectively, which are the most important parameters in understanding viral dynamics. In comparison with previous studies, our population estimate of $c$ (4.736) is greater than the mean estimate of $c$, 3.07 in Perelson et al. (1996) and 3.1 in Perelson and Nelson (1999), further, our population estimate of $c$, 4.736, with CI being (1.570, 8.408) is greater than the population estimate of

TABLE 1

*The summary of estimates of individual dynamic parameters, where SD and CV = SD/Mean denote the standard deviation and coefficient of variation, respectively*

| | $\phi_i$ | $c_i$ | $\delta_i$ | $d_{Ti}$ | $\rho_i$ | $R_{0i}$ |
|---|---|---|---|---|---|---|
| Minimum | 1.216 | 2.157 | 0.080 | 0.002 | 0.208 | 1.043 |
| Median | 10.112 | 4.815 | 0.158 | 0.012 | 1.395 | 2.818 |
| Maximum | 322.851 | 6.448 | 0.341 | 0.196 | 11.319 | 4.064 |
| Mean | 39.755 | 4.827 | 0.170 | 0.033 | 1.472 | 2.726 |
| SD | 69.600 | 0.878 | 0.063 | 0.043 | 2.570 | 0.738 |
| CV(%) | 175.0 | 18.2 | 36.7 | 131.0 | 175.0 | 27.1 |

$c$, 2.81, with CI being (1.24, 6.49) obtained by Han, Chaloner and Perelson (2002) and 3.09 with CI being (2.80, 3.40) in Huang and Wu (2006). Our population estimate of $\delta$ is almost equal to the mean value of $\delta$, 0.37 in Huang and Wu (2006), Stafford et al. (2000) and Klenernam et al. (1996). On the other hand, our population estimate of $\delta$ (0.36) is less than the first-phase decay rate of 0.49 [Perelson et al. (1996)], 0.5 [Perelson and Nelson (1999)] and 0.43 [Nowak et al. (1995)]. In addition, in two separate studies by Perelson et al. (1997) and Markowitz et al. (2003), the mean values of 1.0 and 0.7 for $\delta$ were obtained by holding clearance rate $c$ as constant with values of 23 and 3, respectively, and these two values are substantially greater than our population estimate of 0.36 for $\delta$. Our population estimate of $\rho$ (3.701) is less than those (4.31 and 4.32) for two different data sets estimated by Labbé and Verttoa (2006) and the population estimate of $R_0$ (2.608) is greater than that (1.03) in Labbé and Verttoa (2006). These differences may be due to the various reasons as follows. The analysis of those studies assumed that viral replication was completely stopped by the treatment, they did not incorporate critical clinical factors in the models, and/or they used short-term viral load data to fit their models as well as other issues such as parameter unidentifiability problems in the models. In addition, the first-phase decay rate, estimated from a biexponential viral dynamic model

TABLE 2

*A summary of the estimated posterior means (PM) of population parameters and the corresponding 95% equal-tail credible intervals, where $L_{CI}$ and $R_{CI}$ denote the left and right credible limits of 95% credible intervals*

| | $\phi$ | $c$ | $\delta$ | $d_T$ | $\rho$ | $R_0$ |
|---|---|---|---|---|---|---|
| $PM$ | 14.216 | 4.736 | 0.360 | 0.016 | 3.701 | 2.608 |
| $L_{CI}$ | 3.394 | 1.570 | 0.112 | 0.003 | 0.587 | 1.523 |
| $R_{CI}$ | 37.132 | 8.408 | 0.529 | 0.075 | 7.144 | 4.072 |

[Ho et al. (1995), Perelson et al. (1996), Wu and Ding (1999)] under perfect treatment assumption, is not the exact death rate of infected cells ($\delta$) since the current ARV therapy cannot completely block viral replication [Ding and Wu (1999), Perelson and Nelson (1999), Callaway and Perelson (2002)]. In this study we estimated the death rate of infected cells ($\delta$) directly by accounting for the nonperfect treatment with time-varying drug efficacy. Note that we are unable to validate our results of the other parameter estimates as no conclusive or comparable estimates have been published to date.

As an example, Figure 4 presents six individually fitted curves (solid lines) with observed viral load data in $\log_{10}$ scale as well as the estimated drug efficacy $\hat{\gamma}(t)$ (dotted lines) and the corresponding threshold $e_c$ (broken lines). The constant $\hat{\gamma}(t)$ indicates that both adherence rate $A(t)$ and $IC_{50}(t)$ were held constant over time. It can be seen that the model provides a good fit to the observed data for these subjects. Notably, by comparing the fitted curves and estimated drug efficacy $\hat{\gamma}(t)$, we have seen that, in general, if $\hat{\gamma}(t)$ falls below the threshold $e_c$, viral load rebounds as shown in subjects four and six, and in contrast, if $\hat{\gamma}(t)$ is above $e_c$, the corresponding viral load does not rebound, which are consistent with our theoretical analysis of the dynamic models [Huang, Rosenkranz and Wu (2003)]. It is also important that we can estimate the threshold of the drug efficacy ($e_c$) which is an indicator of drug potency; $e_c$ may reflect the immune response of a patient for controlling virus replications by the patient's immune system and, thus, may pave the way for clinicians to assess whether an ARV regimen is potent enough to suppress viral replication for a patient and to decide whether regimens should be switched. In addition, ARV drugs may show different potencies (Figure 4) for different patients being treated on the same regimen. This may be explained by the fact that a large between-subject variation in estimates of individual parameters may be observed from our modeling approach (see Figure 3), suggesting that parameter estimates such as $\delta$, $c$ may be an important indicator for clinicians to choose individualized ARV therapy.

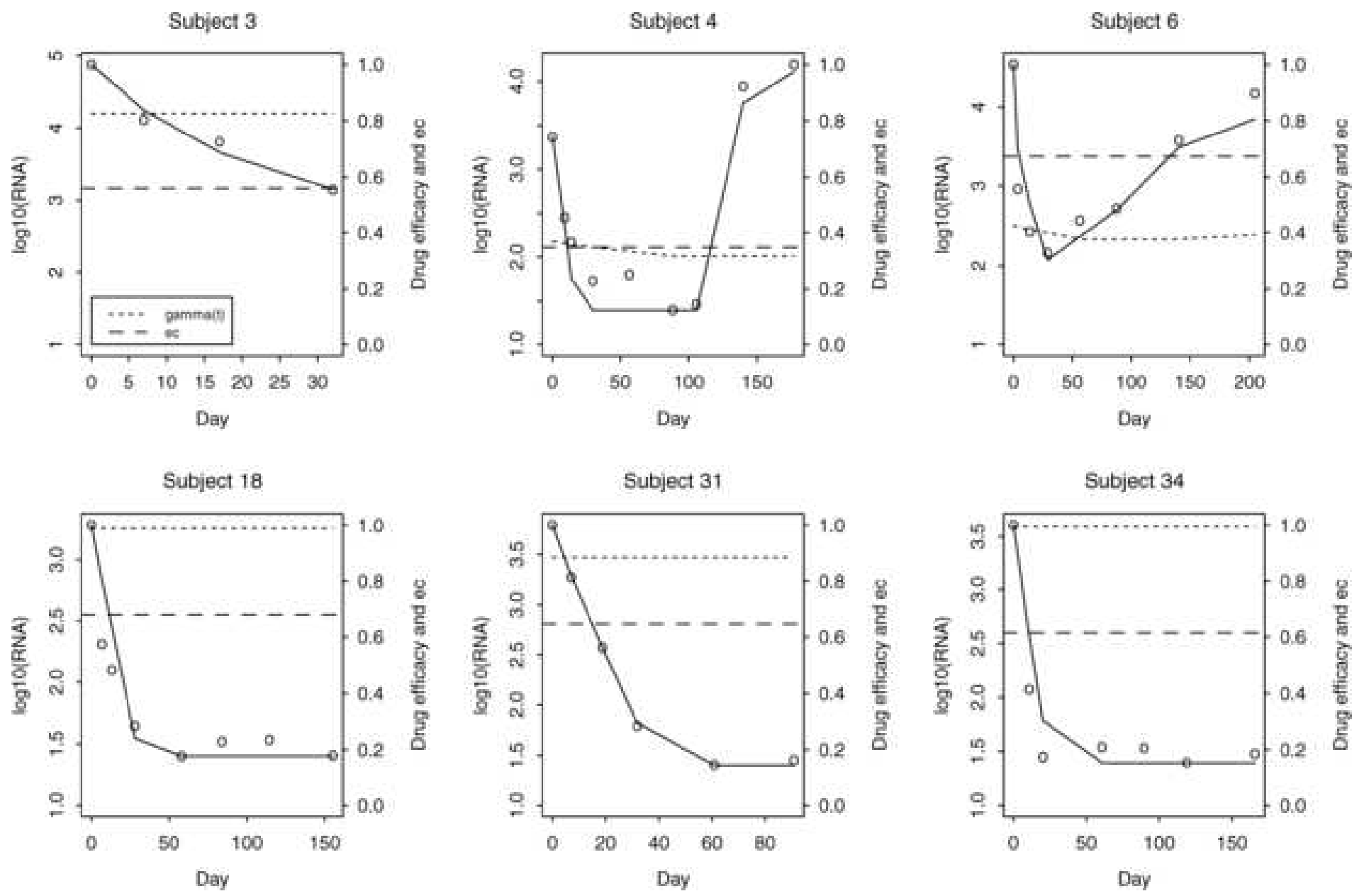

Fig. 4. *Individual fitted curves (solid line) with observed viral load measurements in* $\log_{10}$ *scale, estimated drug efficacies* $\hat{\gamma}(t)$ *(dotted line) and corresponding thresholds* $e_c$ *(broken line) for the six representative subjects.*

4.2. *Correlations between baseline factors and viral dynamic parameters.* We have correlated the baseline factors such as baseline viral load (copies/mL), CD4 cells (cells/mm$^3$), age and weight of patients with the estimated viral dynamic parameters using the Spearman rank correlation test. Baseline viral load and CD4 cell counts were significantly correlated with most of the viral dynamic parameters. These correlations are plotted in Figure 5. No significant correlation was observed between the age or weight of patients and viral dynamic parameters.

Some correlations between baseline viral load and viral dynamic parameters are interesting. A strong negative correlation ($r = -0.727, p < 0.0001$) between baseline viral load and viral clearance rate ($c$) and a positive correlation ($r = 0.711, p < 0.0001$) between baseline viral load and viral load scaling factor ($\rho$) were consistent with the fact that the slower clearance rate of virions and a larger viral load scaling factor, which is the equivalent of a volume of distribution for pharmacokinetics, result in a higher viral load. The positive correlation ($r = 0.694, p < 0.0001$) between baseline viral load and the death rate of target cells ($d_T$) is also interpretable. One possible interpretation is that the higher death rate of target cells may lead to more targets for HIV to infect, which may result in a higher baseline viral load level. A strong positive correlation ($r = 0.761, p < 0.0001$) between baseline viral load and drug efficacy parameter ($\phi$) indicates that the larger value of $\phi$, which corresponds to the lower drug efficacy (because drug efficacy decreases when $\phi$ increases), leads to a higher viral load. The positive correlation ($r = 0.488, p = 0.0018$) between baseline viral load and the death rate of infected cells ($\delta$) may indicate that the more virus (higher baseline viral load) may accelerate the apoptosis of infected cells. This may also clarify the conflicting results in the recent literature on correlation between baseline viral load and first-phase viral decay rate. For example, Notermans et al. (1998) and Wu et al. (2004) reported that plasma baseline HIV-1 RNA levels were positively correlated with first phase viral decay rates; in contrast, Wu et al.

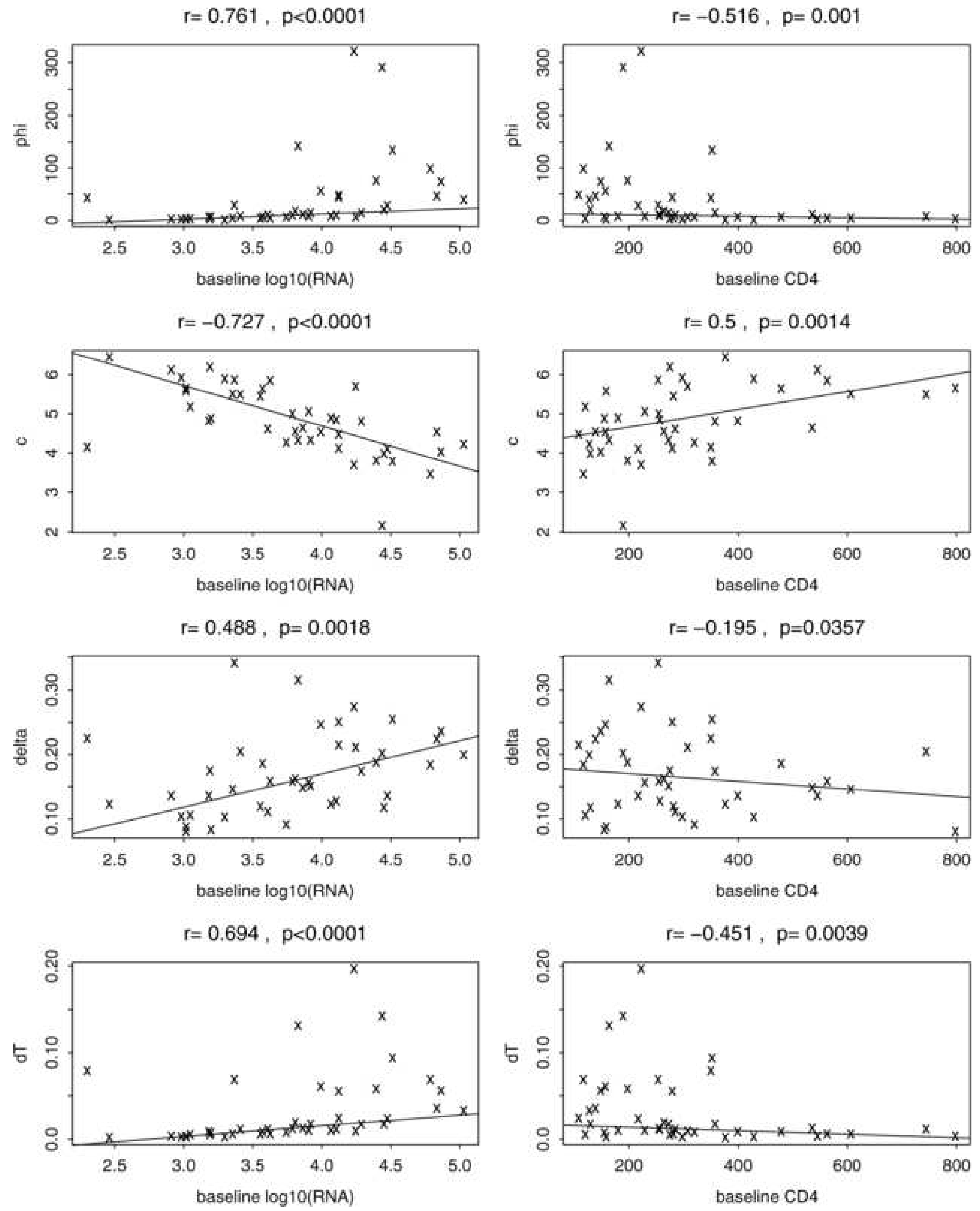

Fig. 5. *Correlations of baseline* $\log_{10}$*(RNA), baseline CD4 cell counts with estimated dynamic parameters. The correlation coefficients and p-values are obtained from the Spearman rank correlation test.*

(1999) and Wu et al. (2003) found a negative correlation between baseline plasma HIV RNA and first phase viral decay rates in two studies. These confused results may be due to the following reason. The first-phase viral decay rate, estimated from a biexponential viral dynamic model [Wu et al.

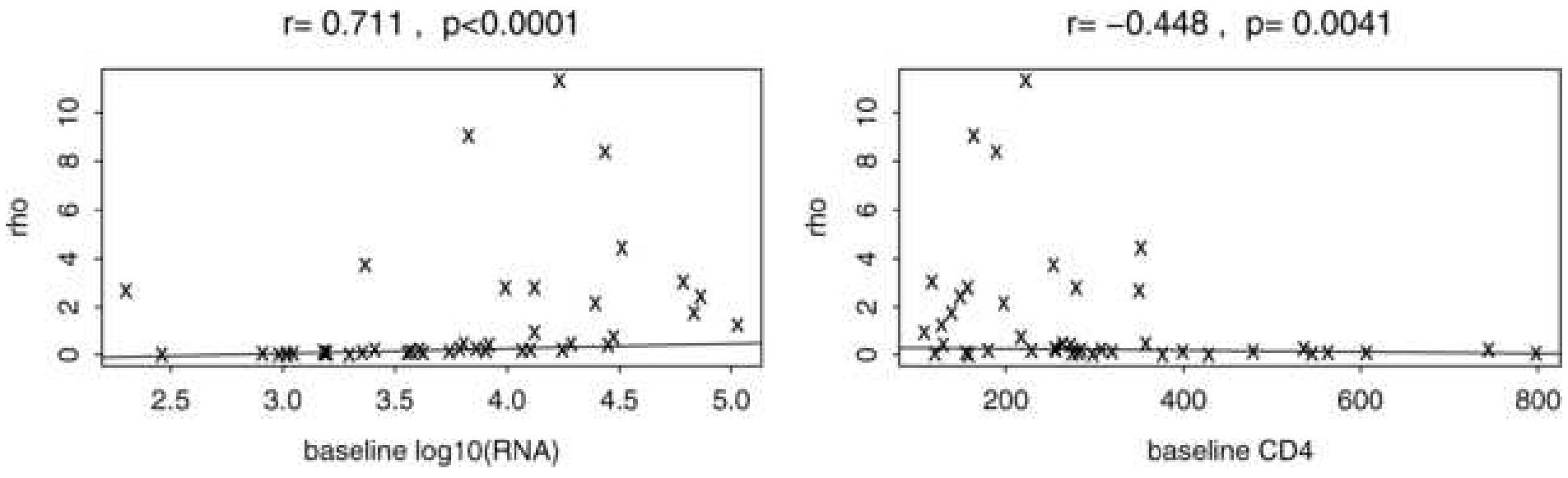

Fig. 5. *Continued.*

(1999), Ding and Wu (2001)] under perfect treatment assumption, is not the true death rate of infected cells ($\delta$) because the current ARV therapy can not completely block viral replication.

It is clearly shown from Figure 5 that baseline $CD4^+$ T cell counts had opposite relationships with the estimated dynamic parameters as baseline viral load had. This is presumably due to a negative correlation between baseline $CD4^+$ $T$ cell count and viral load (data not shown).

4.3. *Relations of viral dynamic parameters with virologic responses.* Virological failure in this study was defined as confirmed two consecutive plasma HIV-1 RNA levels of $\geq$ 200 copies/mL with less than a 1.0-$\log_{10}$ decrease in plasma HIV-1 RNA from baseline by week 8, or failure to achieve a confirmed plasma HIV-1 RNA level of $<$ 200 copies/mL by week 24. Treatment response (success) was defined as two consecutive plasma HIV-1 RNA levels of $<$ 200 copies/mL at any point during the 24-week study [Acosta et al. (2004)].

The estimated viral dynamic parameters in patients with virological success were compared to those in patients with virological failure using the Wilcoxon rank sum test (Figure 6). First we included 33 patients with confirmed virological success or failure status. The patients with virological success had significantly smaller $\phi$, higher clearance rate of free virions ($c$), lower death rate of infected cells ($\delta$), lower death rate of target $T$ cells ($d_T$), smaller viral load scaling factor ($\rho$), and higher basic reproductive ratio for the virus ($R_0$). Intent-to-treat analysis (treating missing data cases as failure) produced similar results.

We correlated the estimated viral dynamic parameters with the viral load changes from baseline to weeks 2, 4 (short-term) and 24 (longer-term). We found that some of the dynamic parameters were significantly correlated with weeks 2 and 4 (short-term) virological responses, but not week 24 (long-term) virological responses. The death rate of infected cells ($\delta$) was positively correlated with short-term weeks 2 and 4 viral load reduction in $\log_{10}$ scale ($r = 0.35$ with $p = 0.029$ at week 2 and $r = 0.38$ with $p = 0.019$ at week 4).

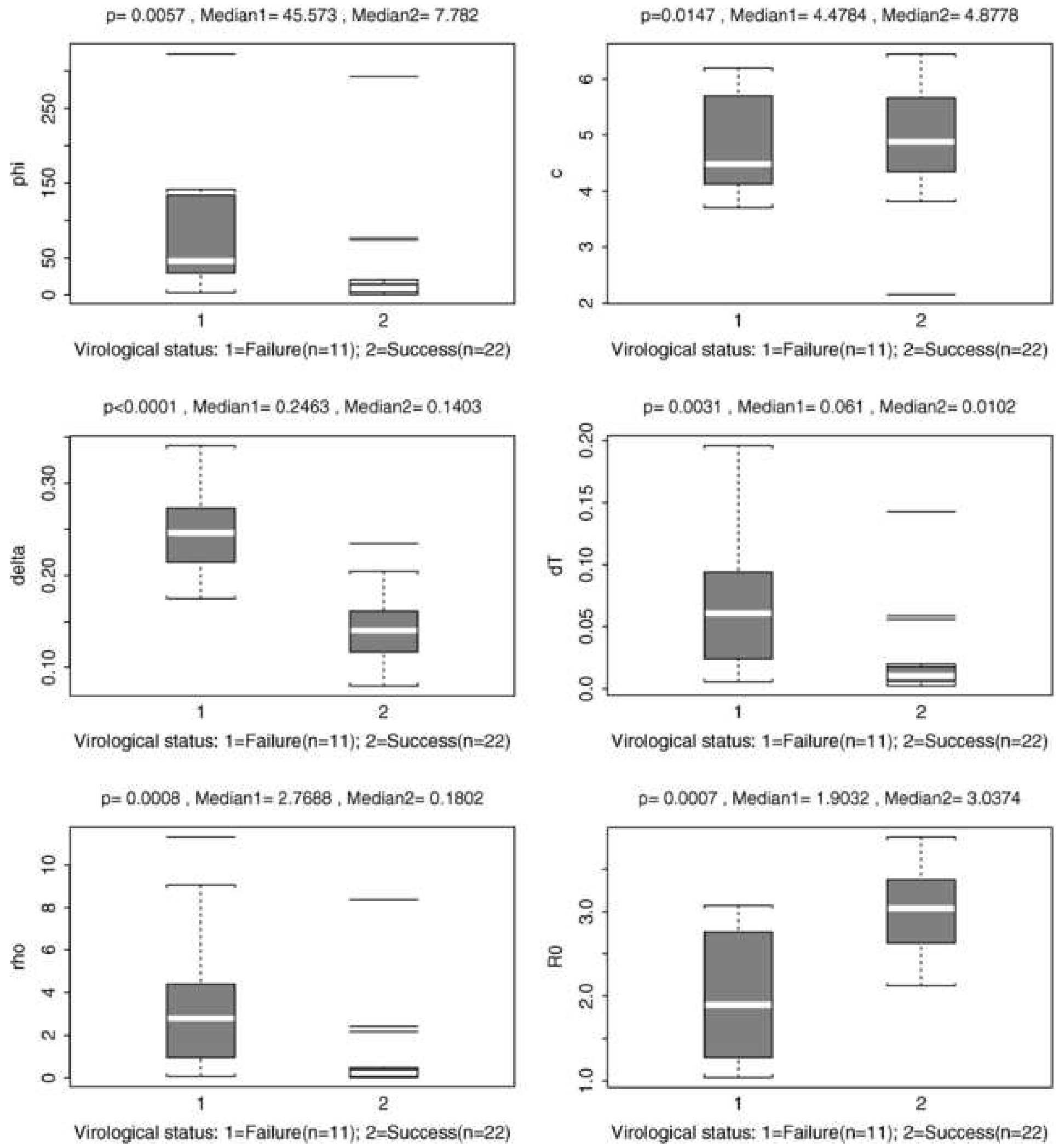

Fig. 6. *Comparisons of estimated dynamic parameters for patients with virological success versus failure (9 patients were excluded due to missing virological response data).*

The death rate of target cells ($d_T$) was positively correlated with short-term weeks 2 ($r = 0.45$ with $p = 0.0044$) and 4 ($r = 0.40$ with $p = 0.015$). These correlations suggest that the higher death rates of infected cells and target cells may result in a larger viral load decline at short-term from baseline.

Based on the protocol, the patient's virological response status was classified as success, failure and missing for a longer-term (24 weeks) response. The patients with virological success had a significantly smaller value of $\phi$, which indicates a stronger drug efficacy (more potent regimen) obtained. It is also understandable biologically that the patients with virological success had a significantly higher clearance rate of free virions ($c$), lower death rate of target $T$ cells ($d_T$) and a smaller value of viral load scaling factor ($\rho$). The patients with virological success had a significantly lower death rate of infected cells ($\delta$) and a higher basic reproductive ratio ($R_0$); and these results were not always consistent with the analysis results of short-term response correlations, which may suggest a discrepancy between the short-term viral response and the longer-term response. For example, the higher death rate of infected cells ($\delta$) may result in a larger viral load reduction in a short-term (2 to 4 weeks), but may not increase the likelihood of virological success in the longer-term (24 weeks). Although this is perplexing, if immune responses are important in virological success, perhaps such responses are continued if infected cells live longer.

**5. Discussion.** The advent of HAART has provided a wealth of information on the interaction between HIV and the human immune system, and is continuing to stimulate the debate on the basic mechanism of viral pathogenesis. However, most of the models developed by biomathematicians and biologists are too complicated and contain too many unknown parameters to be used to analyze real clinical data [Wein, Damato and Perelson (1998), Nowak and May (2000), Stafford et al. (2000)]. In the past decade, some simplified models have been proposed and applied to real viral load data [Ho et al. (1995), Perelson et al. (1996, 1997), Wu et al. (1999), Ding and Wu (2000), Markowitz et al. (2003)]. However, most of these studies only modeled viral dynamic data in a short time period (2 to 8 weeks) after initiating an ARV treatment which was frequently assumed to be 100% effective. In this article we developed a mechanism-based nonlinear time-varying differential equation model for characterizing long-term dynamics to (i) establish the relationship of virological response (viral load trajectory) with drug exposure (pharmacokinetics and adherence) and drug resistance ($IC_{50}$), (ii) to describe both suppression and resurgence of virus, (iii) to directly incorporate observed drug concentration, adherence and drug susceptibility into a function of treatment efficacy and (iv) to use a hierarchical BNLME modeling approach that can not only combine prior information with current clinical data for estimating dynamic parameters, but also characterize inter-subject variability. Thus, the results of estimated dynamic parameters based on this model should be more reliable and reasonable to interpret long-term HIV dynamics. Our models are simplified with the main goals of retaining crucial features of HIV dynamics and, at the same time, guarantying their applicability to typical clinical data, in particular, long-term viral load measurements. We investigated a hierarchical BNLME modeling approach to estimate dynamic parameters in the reparametrized model (2.7) for long-term HIV dynamics. The proposed model fitted the clinical data reasonably well for most patients in our study, although the fitting for a few patients (less than 10%) was not completely satisfactory due to unusual viral load fluctuation patterns, inaccurate measurements of drug exposure and/or adherence for these subjects. For example, self-reported pill count measurements may not reliably reflect actual adherence profiles for some subjects.

We investigated the correlations between baseline factors and estimated dynamic parameters. Some biologically meaningful and interesting correlation results were found. Thus, we may be able to use the baseline viral load or CD4 count to determine whether a treatment regimen is potent enough for a particular patient, as dynamic parameters such as the death rate of infected cells ($\delta$) or viral clearance rate ($c$) can be an indicator of treatment potency. These correlations will help clinicians to select a treatment for their patients. The estimated dynamic parameters in patients with virologic success were compared to those in patients with virologic failure and significantly important findings were summarized in Section 4.

The basic tool for investigating model uncertainty is the sensitivity analysis. That is, we simply make reasonable modifications to the assumptions in question, recompute the posterior quantities of interest, and see whether they have changed in a way that significantly affects the resulting interpretations or conclusions. If the results are robust against the suspected assumptions, we can report the results with confidence and our conclusions will be solid. However, if the results are sensitive to the assumptions, we choose to communicate the sensitivity results and interpret the results with caution [Gamerman (1997)]. In order to examine the dependence of dynamic parameter estimates on the prior distributions and initial values, we conducted the sensitivity analyses using the different mean vector $\boldsymbol{\eta}$ of prior distributions and different initial values (data not shown). The sensitivity analysis results can be summarized as follows: (i) The estimated dynamic parameters were not sensitive to both priors and/or the initial values, and final results are reasonable and robust. (ii) When different priors and/or different initials were used, the results follow the same patterns as those presented in this paper. The conclusions of our analysis remain unchanged.

Although the analysis presented here used a simplified model which appeared to perform well in capturing and explaining the observed patterns, and characterizing the biological mechanisms of HIV infection under relatively complex clinical situations, some limitations exist for the proposed modeling method. First, our model is a simplified model and there are many possible variations [Perelson and Nelson (1999), Nowak and May (2000), Callaway and Perelson (2002)]. We did not separately consider the compartments of short-lived productively infected cells, long-lived and latently infected cells [Perelson et al. (1997)]. Instead, we examined a pooled productively infected cell population. The virus compartment was not further decomposed into infectious virions and noninfectious virions as in Perelson et al.

(1996). Thus, different mechanisms of NRTI and PI drug effects were not modeled. In fact, we only considered PI drug effects in the drug efficacy model (2.6) since the information of NRTI drugs was not collected in our study and the effect of NRTI drugs was considered less important compared to the PI drugs. Second, one may notice that we only have the $IC_{50}$ data at baseline and failure time. We extrapolated the $IC_{50}$ data linearly to the whole treatment period in our modeling. The linear extrapolation is the best approximation that we can get from the sparse $IC_{50}$ data [Wu et al. (2005)]. The linear assumption might have some influence on the estimation results since the $IC_{50}$ might have jumped to a higher level earlier before the failure time when we obtained the sample for the drug resistance test. But we expect that this assumption had little effect on the prediction of virological response since we had relatively frequent monitoring (monthly in the later stage) of virological failure in this study. Third, as measurements of adherence, pill counts, may not reflect actual adherence profiles for individual patients, the data quality would affect our estimation results for viral dynamic parameters. More accurate measurements for adherence such as electronic monitoring devices (e.g., MEMs caps) may improve data quality. Further studies on these issues are definitely needed. Nevertheless, these limitations would not offset the major findings from our modeling approach, although further improvement may be warranted.

We assumed that the distribution of the random effects $\mathbf{b}_i$ is normal. However, due to the nature of AIDS clinical data, it is possible that the data may contain outlying individuals and, thus, may result in a skewed distribution of individual parameters, that is, the random effects may not follow a normal distribution. As Wakefield (1996) suggested, a $t$ distribution may be used which is more robust to outlying individuals than the normal distribution. An extended direction is to incorporate baseline characteristics, such as age and weight, in our drug efficacy model to test the effect of these covariates on each parameter. In addition, this paper combined new technologies in mathematical modeling and statistical inference with advances in HIV/AIDS dynamics and ARV treatment to quantify complex HIV disease mechanisms. The complex nature of HIV/AIDS ARV therapy will naturally pose some challenges, including missing data and measurement error in covariates. These complicated problems, which are beyond the purpose of this article, may be addressed using two-step methods [Higgins, Davidian and Giltinan (1996)] and the joint model method [Wu (2002)]. We are actively investigating these problems now. We hope that we could report these interesting results in the near future.

In summary, the mechanism-based dynamic model is powerful and efficient to characterize relations between antiviral response and drug exposures, drug susceptibility, although some biological assumptions are required. The fitting of a model specified as a set of nonlinear differential equations is routinely done in many fields (in particular, pharmacokinetics and pharmacodynamics, which are closely associated with the analysis of clinical data considered in this article). Long-term viral dynamics can be reasonably modeled with careful considerations of the effects of pharmacokinetics, adherence and drug resistance. Dynamic parameters for individual subjects can be estimated by borrowing information from prior population estimates and across subjects in the same patient population using the novel Bayesian approach. The established models may also be used to simulate antiviral responses of new antiviral agents and will have a notable impact on our understanding of HIV pathogenesis, long-term care with HIV-infected patients and design of new treatment strategies. Although this paper concentrated on HIV dynamics, the basic concept of longitudinal dynamic systems and the proposed methodologies in this study are generally applicable to dynamic systems in other fields such as biology, biomedicine, PK/PD studies or physics, as long as they meet the relevant technical specification (e.g., specified by a set of differential equations).

**Acknowledgments.** The authors thank the Editor and an Associate Editor for their insightful comments and suggestions that led to a marked improvement of the article. We are also indebted to Drs. John G. Gerber, Edward P. Acosta and other A5055 study investigators for their collaborations and allowing us to use the clinical data from their study.

Department of Epidemiology and Biostatistics
College of Public Health, MDC 56
University of South Florida
Tampa, Florida 33612
USA
E-mail: yhuang@health.usf.edu
tlu2002@hotmail.com